\documentclass[twocolumn,showpacs,prb]{revtex4}
\usepackage{subfigure,color}
\usepackage[dvipdfm]{graphicx}
\usepackage{dcolumn}
\usepackage{amsmath}
\usepackage{amssymb}

\makeatletter
\def\btt#1{\texttt{\@backslashchar#1}}
\DeclareRobustCommand\bblash{\btt{\@backslashchar}} \makeatother

\begin{document}

\title{Interface Metallic States between a Topological Insulator and a Ferromagnetic Insulator}

\author{Tetsuro Habe$^1$ and Yasuhiro Asano$^{1,2}$}
\affiliation{$^1$Department of Applied Physics,
Hokkaido University, Sapporo 060-8628, Japan}
\affiliation{$^2$Center for Topological Science \& Technology,
Hokkaido University, Sapporo 060-8628, Japan}

\date{\today}

\begin{abstract}
We study electronic structures at an interface between a topological insulator and a ferromagnetic insulator 
by using three-dimensional two-band model. 
In usual ferromagnetic insulators, the exchange potential is much larger than the bulk gap size in 
the topological insulators and electronic structures are asymmetric with respect to the fermi level.
In such situation, we show that unusual metallic states appear under
the magnetic moment pointing the perpendicular direction to the junction plane, which cannot be 
described by the two-dimensional effective model around the Dirac point. 
 When the magnetic moment is in the parallel direction to the plane, 
the number of Dirac cones becomes even integers. 
The conclusions obtained in analytical calculations are confirmed by numerical simulations 
on tight-binding lattice.
\end{abstract}


\maketitle

\section{introduction}
Physics of a metallic state on a surface of a three-dimensional (3D) topological 
insulator (TI)\cite{Fu2007,Fu2007-2,Moore2007,Chen2009}
is undoubtedly a hot issue these days\cite{Hasan2010,Murakami2011,Qi2011}.
Intrinsic phenomena originated from the topological nature of the insulating state 
would open a novel field of condensed matter physics. 
In particular, the metallic surface state shows interesting features when the TI is attached 
to another materials with gapped excitation spectra such as 
superconductors\cite{Fu2008,Stanescu2010,Lababidi2011} and ferromagnetic insulators\cite{Tanaka2009,Burkov2010,Yokoyama2010,Yokoyama2010-2,Garate2010,Nomura2010}.
The existence of Majorana fermion has been discussed in hybrid structures of such 
materials.\cite{Fu2008,Tanaka2009,Akhmerov2009,Linder2010,Chen2010}

The surface metallic state has a linear dispersion so called two-dimensional 
Dirac cone. The upper and lower corns touch at a point in Brillouin zone, 
so called Dirac point.
The 3D TI's can be classified into the strong and week TI in terms of 
the number of Dirac points\cite{Fu2007,Moore2007,Roushan2009,T.Zhang2009}.
Namely the metallic state is protected from the impurity scattering for 
odd number Dirac cones\cite{Roushan2009,He2011}, whereas it disappears for even number Dirac cones. 
To discuss intrinsic phenomena of TI's, it is necessary to tune the Fermi level
near the Dirac point, which is possible in experiments by chemical 
doping\cite{Hsieh2009,Zhang2010}.

At the interface of a TI and a ferromagnetic insulator (FI),
the metallic state is drastically modified depending on the direction 
of magnetic moment\cite{Tanaka2009}.
The metallic state becomes insulating in the presence of magnetic moment
perpendicular to the interface plane. 
On the other hand, it remains metallic in the presence of 
magnetic moment parallel to the interface. 
The parallel magnetic moments only shift the Dirac point from the $\Gamma$ point
in the Brillouin zone to another points there. 
Such conclusions have been obtained by analyzing effective theoretical model around the Dirac point, 
where the surface state is described by the two-dimensional Dirac Hamiltonian 
under the small exchange potential due to the magnetic moment. 
However it is unclear if 
these conclusions are still valid or not in real TI/FI junctions
because the exchange potential of FI is much larger than the gap of TI.

In this paper, we study electronic states
 at the interface of FI/TI junction by using three-dimensional 
 two-band model.
We show that asymmetry of the band structure in FI
with respect to the Fermi level 
separates the dispersion of interface state from bulk band 
in whole Brillouin zone. This suggests that the effective 
theory around the Dirac point is no longer valid in real TI/FI 
junctions.
A metallic interface state appears even when the magnetic moment in FI is 
perpendicular to the junction plane. 
When the magnetic moment in FI is parallel 
to the interface plane, number of Dirac points should 
be even number in whole Brillouin zone.
In addition to large asymmetry of band structure in FI, 
breaking down the time-reversal symmetry and a basic feature of Brillouin zone
also play important roles in there electric properties of the interface state.
The conclusions obtained by analytical calculation are confirmed by numerical 
simulation on three-dimensional two-band tight-binding lattice.
Obtained results would be 
important not only in the basic physics but also 
in the view of potential application. 

This paper is organized as follows.
 In Sec.~II, we summarize electric property at a TI/FI junction interface 
 based on the effective Hamiltonian around the Dirac point. At the same time,
we discuss the limits of the effective theory.  
In Sec.~III, we analytically study effects of large band asymmetry and large magnetic moment of FI on 
the interface electric states.
In Sec.~IV, the conclusions based on the analytical results are checked by 
the numerical simulation
on three-dimensional tight-binding model.
The conclusion is given in Sec.~V.

\section{Effective theory around the Dirac point}
\begin{figure*}[tbp]
 \begin{center}
   \subfigure{\includegraphics*[height=50mm]{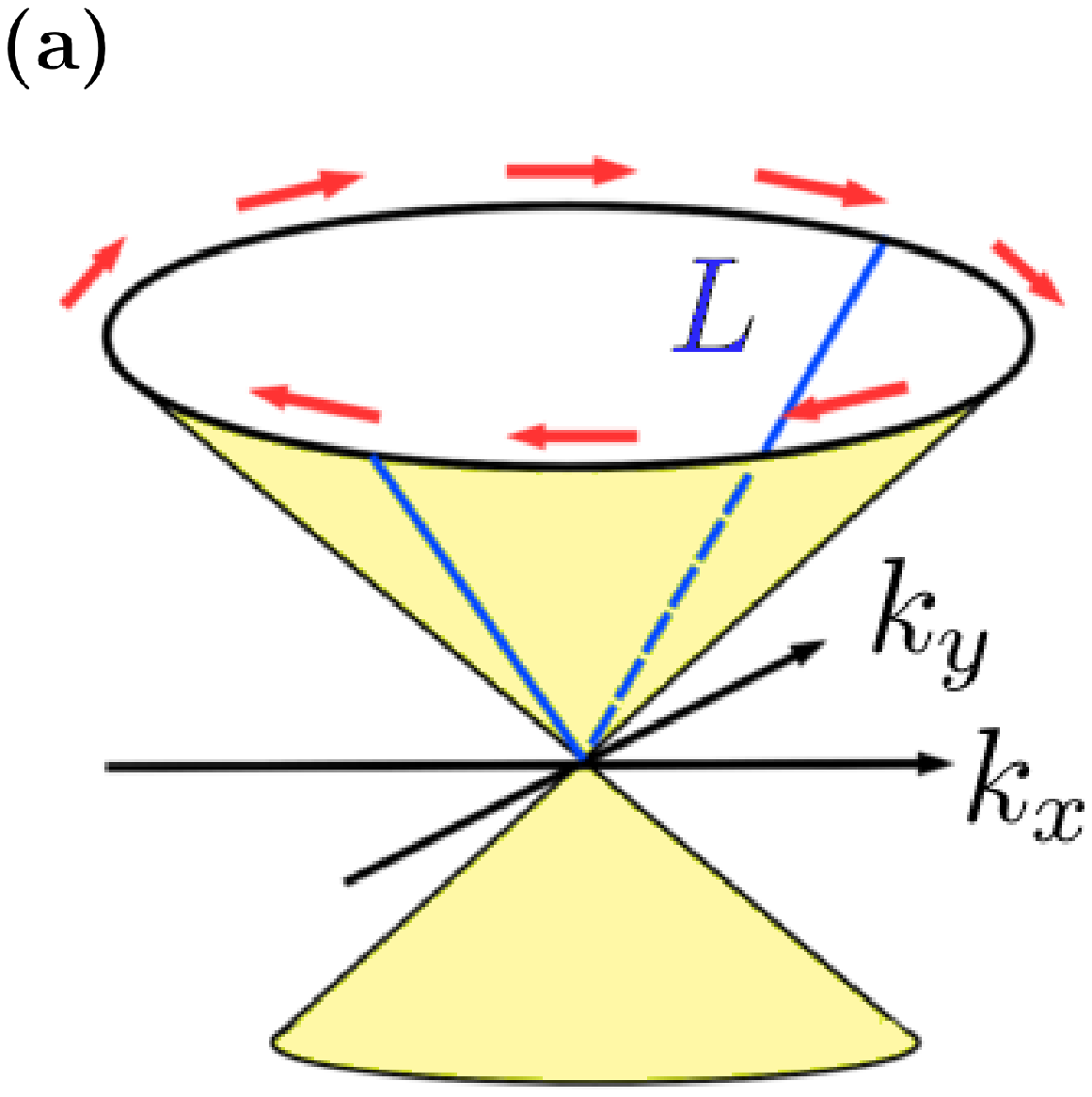}\label{fig:DC}}
 \subfigure{\includegraphics*[height=50mm]{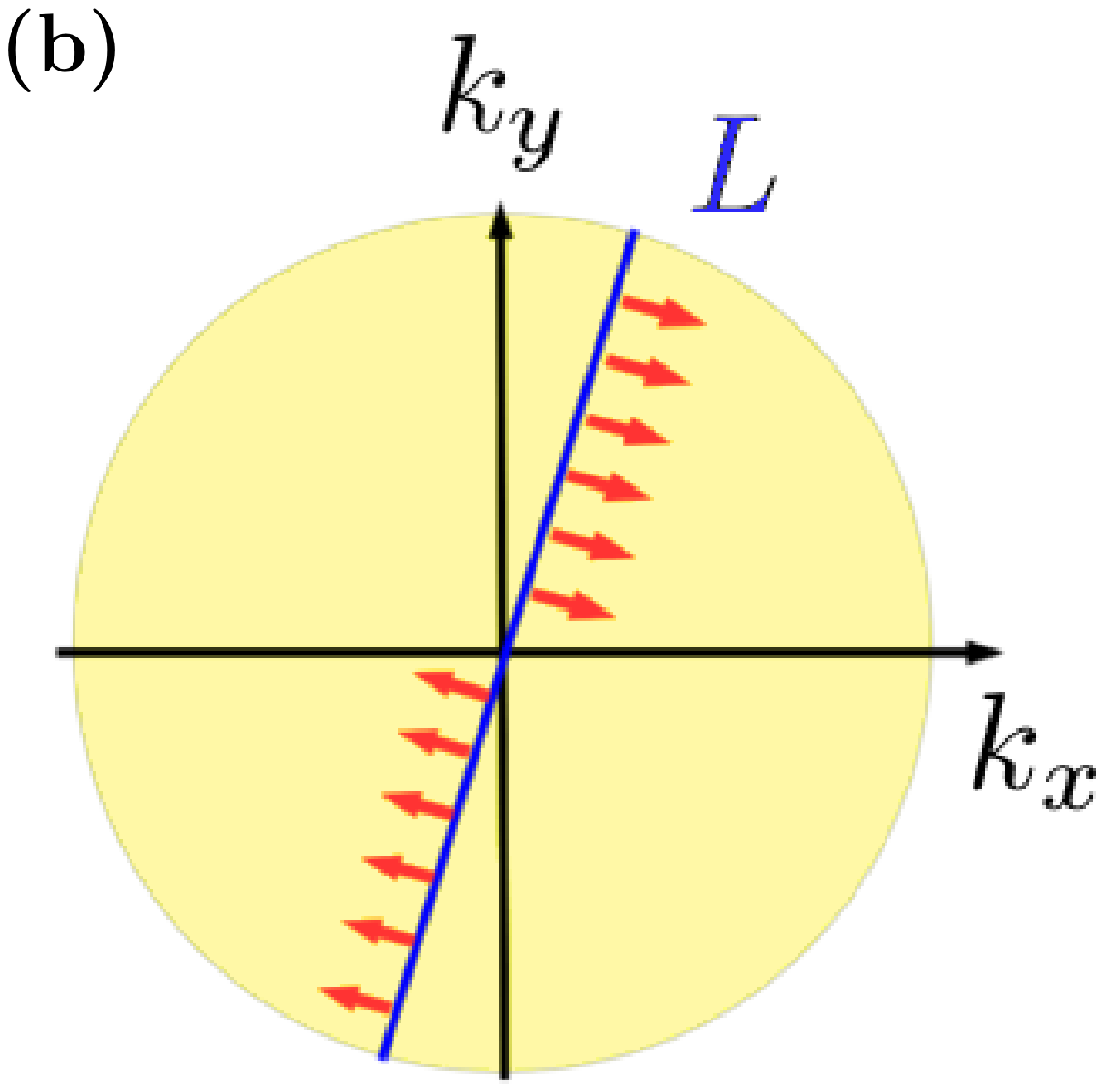}\label{fig:DL}}
  \subfigure{\includegraphics*[height=50mm]{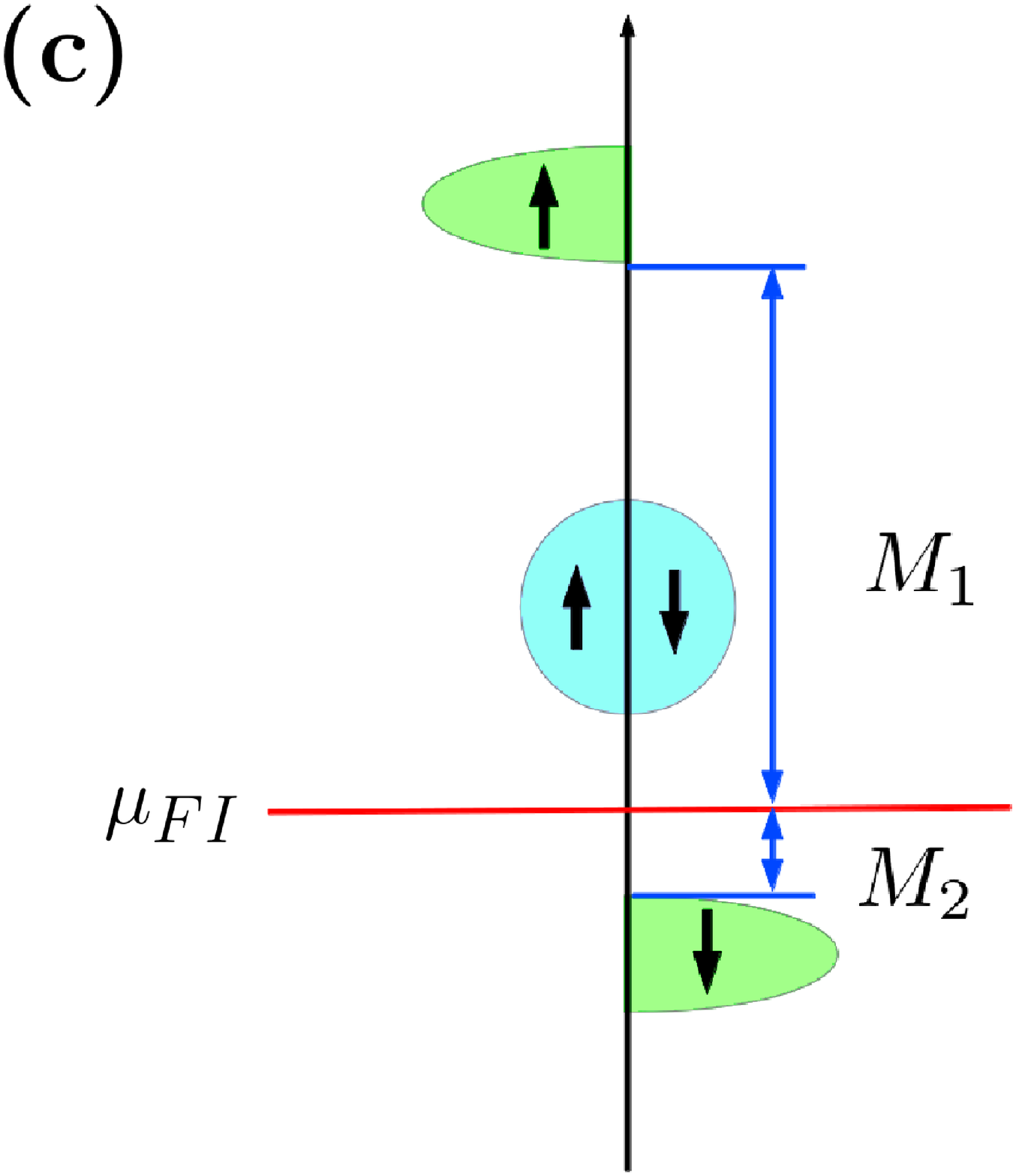}\label{fig:Ferro}}
 \end{center}
\caption{(a) The spin configuration of the Fermi surface.
(b) The spin configuration on the line $L$.
(c) The schematic band structure of a ferromagnetic insulator. 
The arrow in a band denotes spin direction and the horizontal line means the fermi energy.
}
\end{figure*}

We firstly summarize the features of the interface state 
which have been discussed by using effective Hamiltonian
around the Dirac point in two-dimension~\cite{Fu2007,Moore2007}.
The effective Hamiltonian in two-dimension is derived 
from the three-dimensional electric states of a TI described by
\begin{align}
H=&\begin{pmatrix}
A\hat{s}_0&\boldsymbol{d}(\boldsymbol{k})\cdot \hat{\boldsymbol{s}}\\
\boldsymbol{d}(\boldsymbol{k})\cdot\hat{\boldsymbol{s}}&-A \hat{s}_0
\end{pmatrix},\\
A=&M_0-\sum_{\alpha}B_{\alpha}{k_{\alpha}}^2,
\end{align}
where $M_0$ and $B_\alpha$ for $\alpha=1-3$ are band parameters.
The unit matrix in spin space is denoted by $\hat{s}_0$ and
 $\hat{s}_\alpha$ for $\alpha=1-3$ are the Pauli matrices.
The spin-orbit coupling is symbolically expressed by $\boldsymbol{d}
(\boldsymbol{k})$ which satisfies 
\begin{align}
\boldsymbol{d}
(-\boldsymbol{k})=-\boldsymbol{d}
(\boldsymbol{k}).
\end{align}
The surface state on the TI is approximately described by the effective 
Hamiltonian in two-dimension,
 \begin{align}
 h_{\mathrm{sur}}(k_x,k_y)=v_F\boldsymbol{D}\cdot \hat{\boldsymbol{s}}
 -\mu, \label{surface}
 \end{align}
 where $v_F$ is the Fermi velocity.
In what follows, we implicitly consider $\mathrm{Bi_2Se_3}$\cite{Hasan2010}. 
However the arguments below are valid for all TI's.
For $\mathrm{Bi_2Se_3}$, it is shown that $\boldsymbol{D}=(-k_y, k_x)$~\cite{Liu2010}.
The Dirac point is at $(k_x,k_y)=(0,0)$ which we call $\Lambda_0$ in this paper.
The dispersion relation becomes $E_{\boldsymbol{k}}=v_F|\boldsymbol{k}|-\mu$.
The spin configuration on the Fermi surface is schematically illustrated in Fig.\ref{fig:DC}, 
where we assume $\mu>0$ and focus only on the upper Dirac cone. 
The direction of spin and that of momentum are locked to each other. 
Thus the spin direction flips abruptly at $\Lambda_0$ when we trace the 
electronic states along the line $L$ as shown
in Fig.\ref{fig:DL} . Thus the Dirac point may be a kink for the spin polarization 
on a line passing through it. This fact limits the validity of the effective 
theory around the Dirac point.
Namely it is impossible to extend 
the effective theory to electric states in whole Brillouin zone.
Let us trace electronic states along the straight line between $\Lambda_1=(\pi,0)$ 
and $\Lambda_1'=(-\pi,0)$ in the upper Dirac cone.
The states at $\Lambda_1$ and that at $\Lambda_1'$ must be identical to each
other because the two points are connected by a reciprocal vector.
In other words, the topology of the Brillouin zone is the same as that 
of two-dimensional torus ($T^2=S^1\times S^1$).
Although the energy of the two states are equal to each other,
the spin direction of the two states are opposite to each other. 
In the effective theory, $\Lambda_1$ and $\Lambda_1'$ 
characterize the different electronic states.
In real TI's, the effective theory
usually works well because electric states on the Dirac cone is absorbed into the 
bulk energy 
bands before $|\boldsymbol{k}|$ reaching at the zone boundary.

The interface state between a TI and a FI is also approximately
described by the effective Hamiltonian around the Dirac point in two-dimension, 
 \begin{equation}
h_{TIFI}(k_x,k_y) = 
 h_{\mathrm{sur}}(k_x,k_y) 
+\boldsymbol{M}\cdot\hat{\boldsymbol{s}}
\label{sfs2}
 \end{equation}
where 
$\boldsymbol{M}$ is the exchange potential in FI.
Effects of the FI on the interface state are considered only through 
$\boldsymbol{M}$. 
It is easy to show that the magnetic moment 
perpendicular to the two-dimensional plane, $M_z$, 
gives rise to a gap energy at the $\Lambda_0$.
 The magnetic moment parallel to the interface $(M_x,M_y,0)$, 
on the other hand, shifts the Dirac point from $\Lambda_0$ to $(M_y/v_F,-M_x/v_F)$.
In addition to this, the fermi level stays at the Dirac point even 
in the presence of $(M_x,M_y,0)$~\cite{Tanaka2009}.
The conclusions obtained by analyzing Eq.~(\ref{sfs2}) 
seem to be valid for weak exchange potentials smaller than the gap size of TI. 
However, the typical gap size in TI is 100 meV\cite{Chen2009,H.Zhang2009,Hsieh2009-2}, 
whereas the gap of FI is of the order of eV\cite{Borstel1987,Barbagallo2010,Mahadevan2010,An2011}. 
Thus the low energy electronic 
states around the gap of TI should be studied by using more realistic theoretical model.

\section{effects of band asymmetry and large magnetic moment of TI}

Let us consider a TI in three-dimension under the exchange potential 
due to the magnetic moment in a FI.
The Hamiltonian reads
\begin{align}
H=&\begin{pmatrix}
h_0\hat{s}_0&\boldsymbol{d}(\boldsymbol{k})\cdot \hat{\boldsymbol{s}}\\
\boldsymbol{d}(\boldsymbol{k})\cdot \hat{\boldsymbol{s}}&-h_0\hat{s}_0
\end{pmatrix},\label{h:3d}\\
h_0=&M_0-B_1{k_z}^2-B_2({k_x}^2+{k_y}^2),\\
\boldsymbol{d}(\boldsymbol{k})=&(A_2k_x,A_2k_y,A_1k_z),
\end{align}
where $M_0$, $A_1$, $A_2$, $B_1$, and $B_2$ are material parameters\cite{H.Zhang2009}.
The wave number in $x$, $y$ and $z$ directions are denoted by $k_x$, $k_y$ and $k_z$,
respectively.
The Hamiltonian in Eq.~(\ref{h:3d}) is decomposed into two parts
\begin{align}
H=&H_0+H',\\
H_0=&\begin{pmatrix}
(M_0-B_1k_z^2) \hat{s}_0&A_1k_z \hat{s}_z\\
A_1k_z \hat{s}_z&-(M_0-B_1k_z^2) \hat{s}_0
\end{pmatrix},\\
H'=&\begin{pmatrix}
-B_2({k_x}^2+{k_y}^2)\hat{s}_0&A_2(k_x\;\hat{s}_x+k_y\;\hat{s}_y)\\
A_2(k_x\;\hat{s}_x+k_y\;\hat{s}_y)&B_2({k_x}^2+{k_y}^2)\hat{s}_0
\end{pmatrix}.\label{h_kxky}
\end{align}

To analyze interface electric state, we apply the transformation $k_z\rightarrow i\kappa$ in $H_0$,
\begin{align}
H_0=\begin{pmatrix}
(M_0+B_1\kappa^2) \hat{s}_0&iA_1\kappa \hat{s}_z\\
iA_1\kappa \hat{s}_z&-(M_0+B_1\kappa^2) \hat{s}_0
\end{pmatrix}.
\end{align}
In Fig.\ref{fig:Ferro}, schematic band structures of 
Europium chalcogenides are illustrated. 
The band structures are
generally asymmetric with respect to the Fermi level in these materials, 
which we consider through two 
parameters $M_1$ and $M_2$ with $M_1\neq M_2$ as shown in Fig.\ref{fig:Ferro}.
The horizontal line shows the Fermi energy of FI.
The lowest band and the highest one are spin-splitting due to the exchange
potential. We assume that the middle bands are spin-degenerate.
We consider
the large asymmetry of the band structures 
through the 
the exchange Hamiltonian
\begin{align}
H_m=&\begin{pmatrix}
M \hat{s}_{\alpha}+\mu_m \hat{s}_0&0\\
0&0
\end{pmatrix},\\
M_1=& M+\mu_m, \quad M_2=M-\mu_m,
\end{align}
where $\alpha$ indicates the direction of the magnetic moment
in FI
 and $\mu_m$ represents the asymmetry in the band structure.
In these definition, $M_1=M_2$ and $M_1\neq M_2$ describe 
the symmetric and asymmetric band structure, respectively.

\subsection{perpendicular magnetic moment to plane}

When the magnetic moment of FI is perpendicular to the 
junction plane, the exchange Hamiltonian for the surface sate is
\begin{align}
H_m= \begin{pmatrix}
M_1&0&0&0\\
0&-M_2&0&0\\
0&0&0&0\\
0&0&0&0
\end{pmatrix}.
\end{align}
In usual FI's, a relation $M_i\gg M_0$ holds.
The Hamiltonian $H_0+H_m$ is decomposed into two $2\times 2$ 
matrices whose
eigenvalues are $E_i=(\tilde{M}_i-{M}_0)/2$ 
with $\tilde{M}_1=M_1+M_0$ and 
$\tilde{M}_2=-M_2+M_0$.
The eigenstates of can be expressed by
\begin{align}
\boldsymbol{v}_1(\kappa_1)=\begin{pmatrix}
a_1(\kappa_1)\\
0\\
b_1(\kappa_1)\\
0
\end{pmatrix}
,\;\;\;\;\;
\boldsymbol{v}_2(\kappa_2)=\begin{pmatrix}
0\\
a_2(\kappa_2)\\
0\\
b_2(\kappa_2)
\end{pmatrix}.
\end{align}
The coefficients $a_i$ and $b_i$ satisfy
\begin{align}
\frac{a_i}{b_i}
=-\frac{iD_i\kappa}{(\tilde{M}_i+{M}_0)/2-B_1\kappa^2}
\end{align}
where $D_1=A_1$ and $D_2=-A_1$.
This Hamiltonian is equivalent to that of the surface state of a
TI facing to vacuum 
by substituting $(\tilde{M}_i+{M}_0)/2$ by $M_0$.

The imaginary wavenumber $\kappa_i^\pm$ takes 
different forms depending on the sign of $\tilde{M}_i+{M}_0$.
For $\tilde{M}_i+{M}_0>0$, 
$\kappa$ has the similar form as it is in the TI/vacuum surface,
\begin{align}
\kappa_i^\pm=\frac{A_1}{2B_1}\left(1\pm\sqrt{1-\frac{2B_1(\tilde{M}_i+{M}_0)}{{A_1}^2}}\right).\label{kappa1}
\end{align}
The eigenstate in this case can be described by
\begin{align}
\begin{pmatrix}
a_i\\
b_i
\end{pmatrix}
=
\begin{pmatrix}
D_i/A_1\\
i
\end{pmatrix}
\left(C_+e^{-\kappa_i^+z}+C_-e^{-\kappa_i^-z}\right),
\end{align}
with $C_\pm$ being arbitrary constants.
For $\tilde{M}_i+{M}_0<0$, the wavenumber becomes
\begin{align}
\kappa_i^\pm=\frac{A_1}{2B_1}
\left(\sqrt{1-\frac{2B_1(\tilde{M}_i+{M}_0)}{{A_1}^2}}\pm1\right).
\end{align}
The eigenstate is given by
\begin{align}
\begin{pmatrix}
a_i\\
b_i
\end{pmatrix}
=
C_+\begin{pmatrix}
D_i/A_1\\
i
\end{pmatrix}
e^{-\kappa_i^+z}
+C_-
\begin{pmatrix}
-D_i/A_1\\
i
\end{pmatrix}
e^{-\kappa_i^-z}.
\end{align}
For $M_1>0$, $\tilde{M}_1+{M}_0>0$ always holds.  
Thus $\kappa_1$ takes Eq.~(\ref{kappa1}). On the other hand, 
$\tilde{M}_1+{M}_0$ can be either positive or negative
even for $M_2>0$. 

We first analyze weak exchange potential 
satisfying $M_2<2M_0$. 
The wave function of in this case is
\begin{align}
\begin{pmatrix}
a_i\\
b_i
\end{pmatrix}
=
\begin{pmatrix}
D_i/A_1\\
i
\end{pmatrix}
\left(C_i^+\exp[-\kappa_i^+z]+C_i^-\exp[-\kappa_i^-z]\right).
\end{align}
with $C_i^\pm$ being the normalization constant.
For simplicity, in what follows, we drop $z$ dependence 
from the wave function.
There are only two independent wave function for $M_2<2M_0$.
The surface state is a superposition 
of $\psi_1$ and $\psi_2$ which are defined by
\begin{align}
\psi_1=
\frac{1}{\sqrt{2}}
\begin{pmatrix}
1\\
0\\
i\\
0
\end{pmatrix},\;\;\;\;\;
\psi_2=
\frac{1}{\sqrt{2}}
\begin{pmatrix}
0\\
-1\\
0\\
i\\
\end{pmatrix}
.
\end{align}

The total Hamiltonian $H_0+H'+H_m$ can 
be represented in these basis of 
$\psi_i$ as,
\begin{align}
H=&\begin{pmatrix}
M_1&0\\
0&-M_2
\end{pmatrix}
+
\begin{pmatrix}
H'_{11}&H'_{12}\\
H'_{21}&H'_{22}
\end{pmatrix}\nonumber\\
=&
\begin{pmatrix}
M_1&iv_F (k_x-ik_y)\\
-iv_F (k_x+ik_y)&-M_2
\end{pmatrix},\\
H'_{ij}=&\langle{\psi_i}|H'|\psi_j\rangle
\end{align}
with $v_F=A_2$. 
The energy of the surface state is 
\begin{align}
E=\frac{M_1-M_2}{2}\pm\sqrt{\frac{(M_1+M_2)^2}{4}+{v_F}^2{k}^2}
\end{align}
with $k=\sqrt{{k_x}^2+{k_y}^2}$.
For weak exchange potential $M_2<2M_0$,
the exchange potential in the $z$ direction causes the gap,
which is consistent with the previous theories~\cite{Tanaka2009}.
The asymmetry of the band structures gives a constant
energy shift to the dispersion
relation.

Next we consider strong exchange potential 
satisfying $M_2>2M_0$. 
In this case, the straight forward calculation 
of the eigenfunction at the $\Gamma$ point results 
in 
\begin{align}
\psi_1=\frac{1}{\sqrt{2}}\begin{pmatrix}
1\\
0\\
i\\
0
\end{pmatrix}
,\;\;
\psi_2=
\frac{1}{\sqrt{2}}
\begin{pmatrix}
0\\
1\\
0\\
i
\end{pmatrix}
,\;\;
\psi_3=
\frac{1}{\sqrt{2}}
\begin{pmatrix}
0\\
-1\\
0\\
i
\end{pmatrix}.
\end{align}
For convenience, we employ an another basis 
as follows,
\begin{align}
\psi_1'=\frac{1}{\sqrt{2}}\begin{pmatrix}
1\\
0\\
i\\
0
\end{pmatrix}
,\;\;\;\;\;
\psi_2'=
\begin{pmatrix}
0\\
1\\
0\\
0
\end{pmatrix}
,\;\;\;\;\;
\psi_3'=
\begin{pmatrix}
0\\
0\\
0\\
1
\end{pmatrix}.
\end{align}
The total Hamiltonian $H_0+H'+H_m$ in this representation 
reads,
\begin{align}
H=\begin{pmatrix}
M_1&-iv_F(k_x-ik_y)&v_F(k_x-ik_y)\\
iv_F(k_x+ik_y)&-M_2-B_2k^2&0\\
v_F(k_x+ik_y)&0&-M_2+B_2k^2
\end{pmatrix}.
\end{align}
with $v_F=A_2/\sqrt{2}$.
The energy dispersion can be derived from the eigen equation,
\begin{align}
x^3-2Mx^2-({B_2}^2k^4+2{v_F}^2k^2)x+2M{B_2}^2k^4=0,\label{det1}
\end{align}
with $x=E+M_2$.
At the vicinity of $\Gamma$-point, $x(k)$ is approximately given by 
\begin{align}
x(k)=a_0+a_1k^2+a_2k^4.
\end{align}
Here $a_0$ can be obtained easily by putting $k=0$. We obtain two values as
\begin{align}
a_0=0,\;\;2M.
\end{align}
For $a_0=0$, $a_i$ can be derived by putting the coefficients of $k^4$ and $k^6$ terms in Eq.~\ref{det1}
to be zero.
 Since $M \gg M_0$, $a_1$ and $a_2$ have simple expression
\begin{align}
a_1=&-\frac{{v_F}^2}{2M}\pm B_2\sqrt{1+\frac{{v_F}^4}{4M^2{B_2}^2}}\\
\simeq& -\frac{{v_F}^2}{2M}\pm B_2,\\
a_2\simeq&\mp\frac{{v_F}^2}{4M^2}B_2.
\end{align}
Then the energy dispersions are approximately given by
\begin{align}
E(k)=-M_2\pm B_2k^2\mp\frac{{v_F}^2}{4M^2}B_2k^4.
\end{align}
In the same way, 
we also obtain
\begin{align}
E(k)=M_1+\frac{{v_F}^2}{M}k^2-\frac{{v_F}^4}{2M^3}k^4,
\end{align}
for $a_0=2M$.
In both $a_0=0$ and $2M$, the coefficient of $k^2$ and that of 
$k^4$ have opposite sign to each other.
In addition, we can also predict that two minima of the dispersion 
go across the fermi
level and the interface becomes metallic for $M>2M_0$.

\begin{figure*}[htbp]
 \begin{center}
  \subfigure{\includegraphics*[height=50mm]{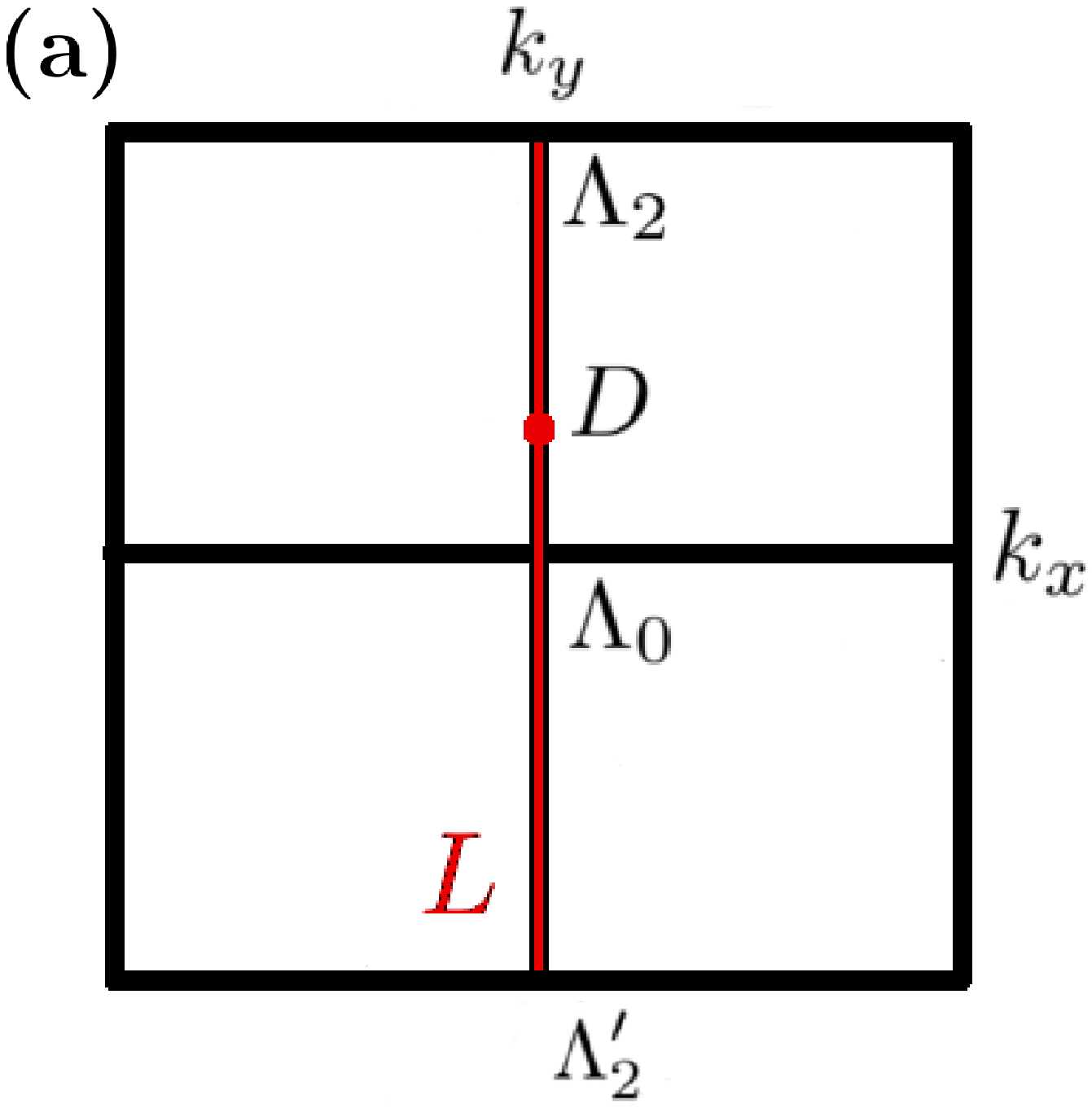}\label{fig:BZ}}
  \subfigure{\includegraphics*[height=50mm]{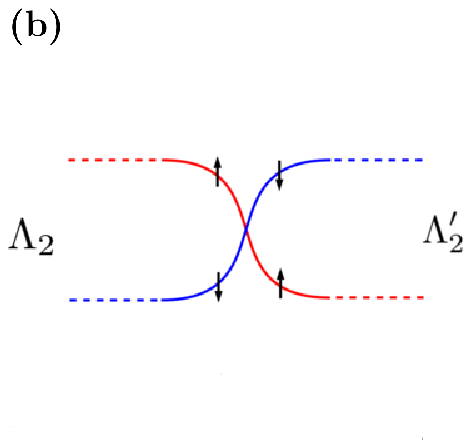}\label{fig:DP}}
 \end{center}
\caption{ (a) The simple Brillouin zone and its TRI points $\Lambda_i$. In this figure, $\Lambda_2$ and $\Lambda_2'$ are same TRI points under translational operation of a reciprocal lattice vector. 
(b) The spin configuration on the line $L$ with a single Dirac points is illustrated.}
 \label{fig:FI}
\end{figure*}

\subsection{parallel magnetic moment to plane}

When the magnetic moment of FI is parallel to the junction plane, 
the Hamiltonian of the surface sate at $\Gamma$-point is
$H_0+H_m$ with
\begin{align}
H_m=&\begin{pmatrix}
M \hat{s}_x+\mu_m\hat{s}_0&0\\
0&0
\end{pmatrix}.
\end{align} 
Here we assume that the magnetic moment is in the $x$ direction. 
This does not loose the generality of argument below because
the Hamiltonian is rotationally invariant in momentum space. 
Applying an unitary transformation,
we obtain
\begin{align}
U^\dagger& (H_0+H_m)U= \nonumber\\
&\begin{pmatrix}
(M_0+B_1\kappa^2)\hat{s}_0+\hat{M}&-iA_1\kappa s_x\\
-iA_1\kappa s_x&-(M_0+B_1\kappa^2)\hat{s}_0
\end{pmatrix},\\
\hat{M}&=\begin{pmatrix}
M_1&0\\
0&-M_2
\end{pmatrix},
\end{align}
with
\begin{align}
U=\begin{pmatrix}
(\hat{s}_0-i\hat{s}_y)/\sqrt{2}&0\\
0&(\hat{s}_0-i\hat{s}_y)/\sqrt{2}
\end{pmatrix}.
\end{align}
The eigenvectors can be expressed by
\begin{align}
\psi_1=\begin{pmatrix}
a_1(\kappa)\\
0\\
0\\
b_1(\kappa)
\end{pmatrix},\;\;\;\;\;
\psi_2=\begin{pmatrix}
0\\
a_2(\kappa)\\
b_2(\kappa)\\
0
\end{pmatrix}.
\end{align}
The elements satisfy
\begin{align}
\frac{
a_i}
{b_i}=\frac{iA_1\kappa}{\tilde{M}_i+B_1\kappa^2}
.
\end{align}
where $\tilde{M}_1=M_0+M$ and $\tilde{M}_2= M_0-M$. 
The eigenvalues and eigenvectors can be calculated in the same way with the previous subsection.

When the exchange potential is weak $M_2<2M_0$, the eigenvectors are
The eigenvalues $E_i$ are
\begin{align}
E_1=\frac{M_1}{2},\;\;\;\;\;E_2=-\frac{M_2}{2}.
\end{align} 
Corresponding vectors are given by
\begin{align}
\psi_1=
\frac{1}{\sqrt{2}}
\begin{pmatrix}
-1\\
0\\
0\\
i
\end{pmatrix},\;\;\;\;\;
\psi_2=
\frac{1}{\sqrt{2}}
\begin{pmatrix}
0\\
-1\\
i\\
0
\end{pmatrix}.
\end{align} 
The total Hamiltonian $H=\tilde{H}_0+U^\dagger H'U$
becomes
\begin{align}
H=\frac{M_1-M_2}{2}s_0+
\begin{pmatrix}
M-v_Fk_y&-iv_Fk_x\\
iv_Fk_x&-M+v_Fk_y
\end{pmatrix},
\end{align}
where $v_F=A_2$ and $2M=M_1+M_2$.
The energy dispersion is given by
\begin{align}
E=\frac{M_1-M_2}{2}\pm v_F\sqrt{{k_x}^2+(k_y-M/v_F)^2}.
\end{align}
The Dirac point moves from the $\Gamma$ point to $(0,M)$, which is consistent with 
the effective theory in around the Dirac point. 
The asymmetry of the band structures, however, shifts the fermi level from the Dirac point.

When the exchange potential is sufficiently large satisfying $M_2>2M_0$, 
the basis of the surface state become 
\begin{align}
\psi_1=\frac{1}{\sqrt{2}}
\begin{pmatrix}
-1\\
0\\
0\\
i
\end{pmatrix},\;\;\;\;\;
\psi_2=
\begin{pmatrix}
0\\
1\\
0\\
0
\end{pmatrix},\;\;\;\;\;
\psi_3=
\begin{pmatrix}
0\\
0\\
1\\
0
\end{pmatrix}.
\end{align}
The total Hamiltonian $H=H_0+H'+H_m$ in this basis
results in
\begin{align}
H=\begin{pmatrix}
M_1-v_Fk_y&-iv_Fk_x/\sqrt{2}&-v_Fk_x/\sqrt{2}\\
iv_Fk_x/\sqrt{2}&-M_2+B_2k^2&iv_Fk_y\\
-v_Fk_x/\sqrt{2}&-iv_Fk_y&-M_2-B_2k^2
\end{pmatrix},\label{H:para}
\end{align}
with $v_F=A_2$.

By analyzing Eq.~(\ref{H:para}) in detail, we can conclude that
(a) there are two Dirac cones in the whole Brillouin zone, (b)
the asymmetry of band structure in FI with respect to the
fermi level may causes the separation of the interface state from the bulk band, 
and
(c) three branches of surface states appear in the gap of TI.
These conclusions can be confirmed in a simple case where we consider the 
 dispersion relation along a line satisfying $k_x=0$.
At $k_x=0$, 
three dispersion branches appear at the interface
\begin{align}
E_1=&M_1-v_Fk_y,\label{e11}\\
E_2=&-M_2- v_Fk_y\sqrt{1+({B_2}^2/{v_F}^2){k_y}^2},\label{e12} \\
E_3=&-M_2+ v_Fk_y\sqrt{1+({B_2}^2/{v_F}^2){k_y}^2}.\label{e13}
\end{align}
Near the $\Gamma$-point, two branches $E_1$ and $E_2$ 
are almost parallel to each other. The remaining branch $E_3$ 
goes across $E_1$ and $E_2$. Therefore there are two Dirac points.
For $k_y'=k_y-(M_1+M_2)/2v_F$, Eq~(\ref{e11}) and (\ref{e13}) can be represented by
\begin{align}
E=&\frac{M_1-M_2}{2}\pm v_Fk_y',
\end{align}
where higher order terms for $k^3$ in Eq~(\ref{e13}) are ignored. 
The first term implies a asymmetry of the band structure of FI. 

As we have discussed above, the asymmetry of band structure in 
FI removes the dispersion 
of the interface state from the bulk band.
 This causes more drastic modification 
of interface state in the presence of magnetic moment parallel to 
the interface plane.
When $\boldsymbol{M}=(M_x,0,0)$, the magnetic moment 
shifts the Dirac point from $\Lambda_0$ in the Brillouin 
zone to a point $D$ as shown in Fig.~\ref{fig:BZ}.
Let us consider the spin configuration along 
a line which satisfies
$\boldsymbol{D} \parallel \boldsymbol{M}$ (Eq. \ref{surface}) and 
passes through the Dirac point $D$.
For $\boldsymbol{M}=(M_x,0,0)$, the line corresponds
to the straight line $L$ connecting $\Lambda_2$ and $\Lambda_2'$
as shown in Fig.~\ref{fig:BZ}.
We note two key features of spin direction along the line:
(i) $\Lambda_2$ and $\Lambda_2'$ are identical 
point to each other and (ii) the spin direction flips at
$D$.
If the number of the Dirac point is one, 
spin direction at $\Lambda_2$ and $\Lambda_2'$ would be 
opposite to each other. This statement, however, contradict 
to (i). Therefore the number of Dirac point must be 
an even integer on $\Lambda_2-\Lambda_2'$.
Since $D$ is a Dirac point, at least one extra Dirac
point is necessary on $\Lambda_2-\Lambda_2'$ (Fig.\ref{fig:DP}).

This conclusion above can be obtained in more general argument.
The Dirac point can be regarded as the magnetic monopole in the momentum space.
The Gauss integration
in the first Brillouin zone becomes finite in the presence of the single
monopole.
This integration should coincide with the path integration of 
$\boldsymbol{D}(\boldsymbol{k})$ along the zone boundary.
However the integration along the boundary vanishes be cause 
of the relation $\boldsymbol{D}(-\boldsymbol{k})=-\boldsymbol{D}(\boldsymbol{k})$.
Thus there must be extra monopoles in the Brillouin zone.
According to this argument, the number of the Dirac points must be 
even number in the Brillouin zone. In Eq.~(\ref{e13}), two Dirac points are 
expected in the present situation.

The conclusions obtained by the analytical calculation are confirmed by
numerical simulation in the next section.

\begin{figure*}[htbp]
 \begin{center}
   \subfigure{\includegraphics*[height=50mm]{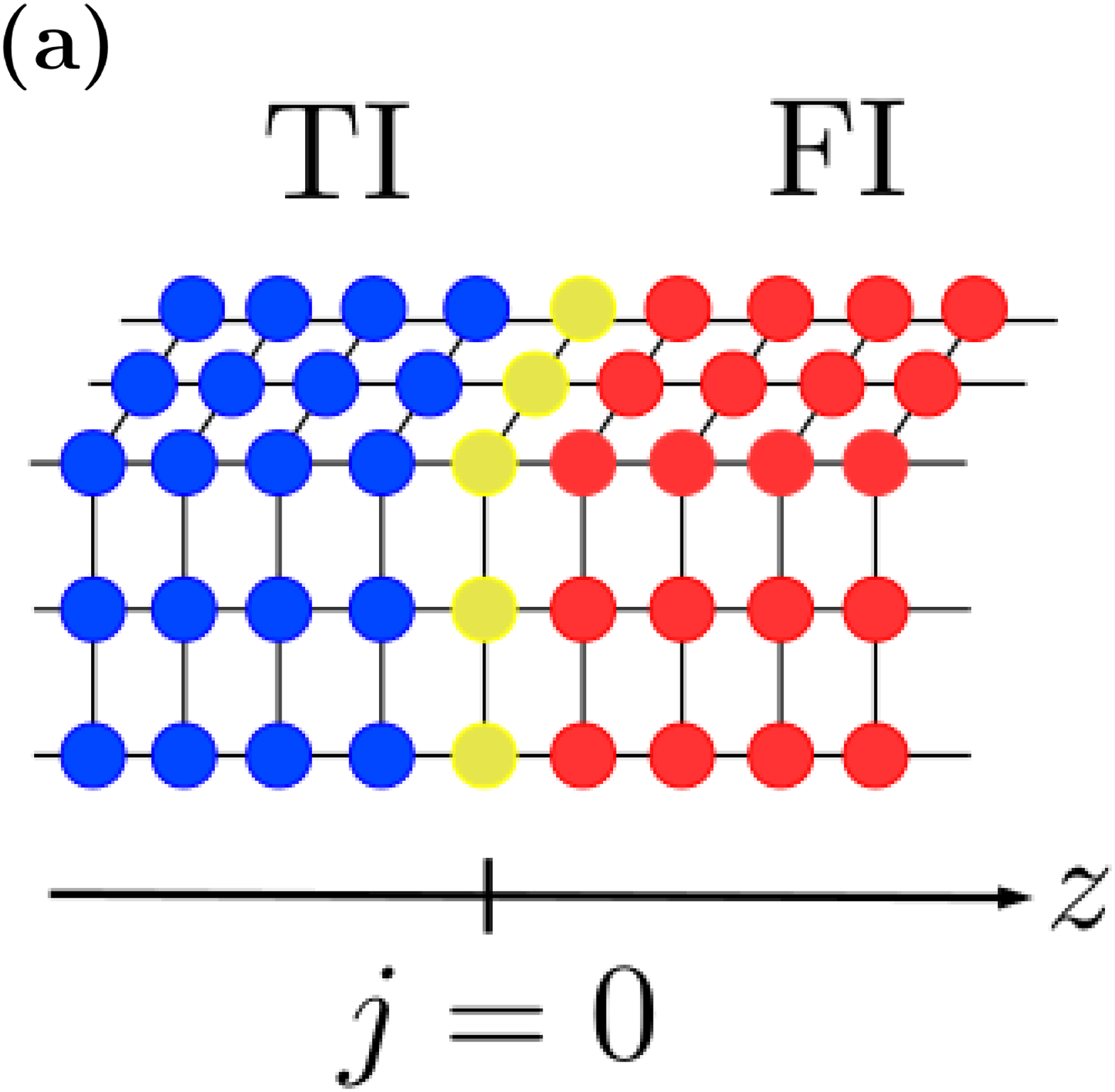}\label{fig:Lattice}}
   \subfigure{\includegraphics*[height=50mm]{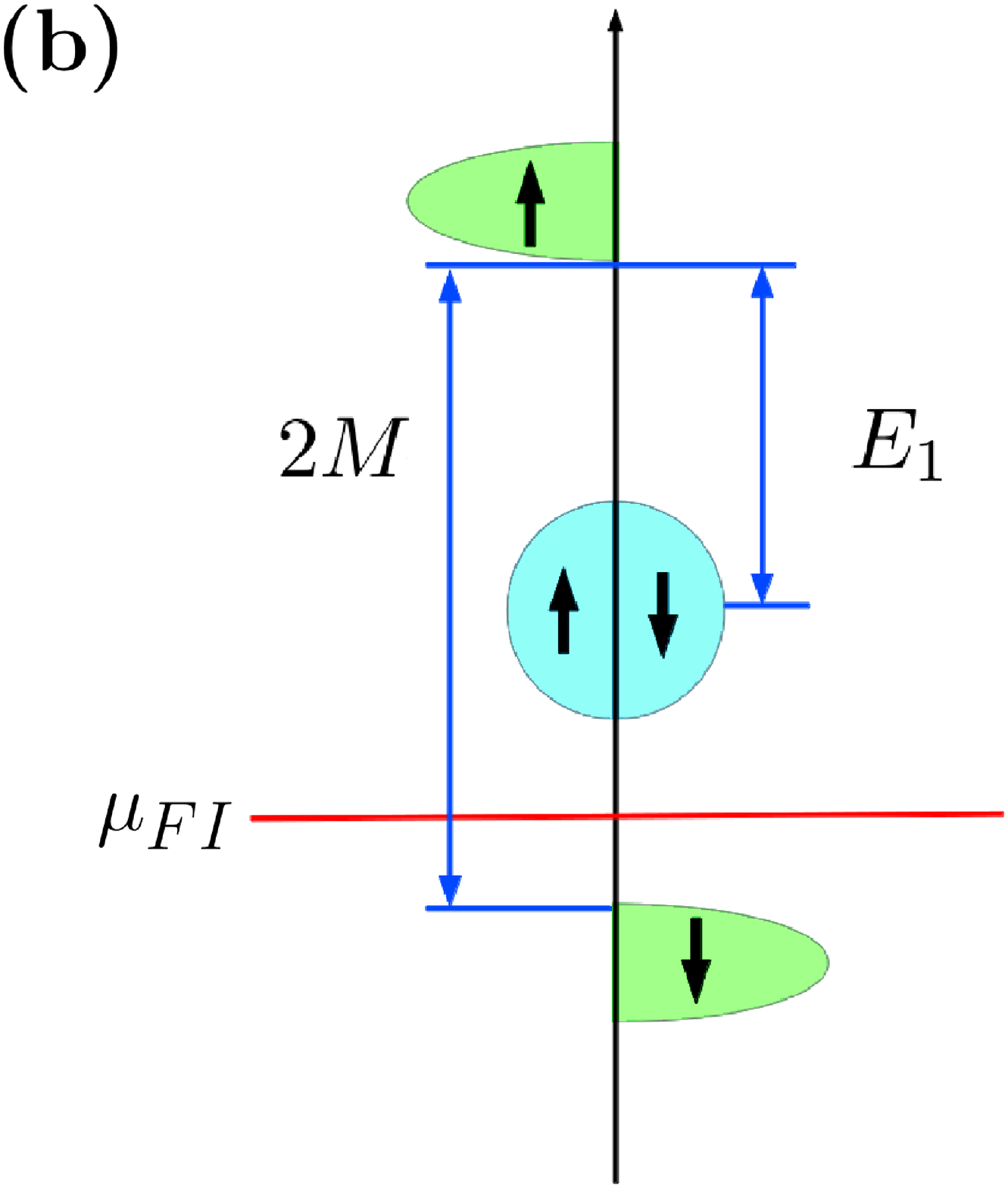}\label{fig:FIscheme}}
 \end{center}
\caption{(a): TI/FI junction on the three-dimensional tight-binding lattice. 
The interface is at $j=0$.
(b): The schematic band structure of a FI. 
The arrow in a band denotes spin direction and the horizontal line labeled by $\epsilon_F$ 
is the fermi energy.}
\end{figure*}

\section{Numerical results in 3D}
Let us consider a junction of TI and FI on three-dimensional tight-binding lattice 
as shown in Fig.~\ref{fig:Lattice}.
We describe the TI by using the two-band model as
\begin{widetext}
\begin{align}
H_{TI} =& \sum_{j,j'}\sum_{\boldsymbol{k}}
\left[
\tilde{c}^\dagger_{\boldsymbol{k},j',1},\tilde{c}^\dagger_{\boldsymbol{k},j',2}\right]
\left[\begin{array}{cc}
\xi_{TI} \hat{s}_0 & \boldsymbol{A}\cdot\hat{\boldsymbol{s}} \\
\boldsymbol{A}\cdot\hat{\boldsymbol{s}} & - \xi_{TI}\hat{s}_0
\end{array}\right]
\left[\begin{array}{c}
\tilde{c}_{\boldsymbol{k},j,1}\\ \tilde{c}_{\boldsymbol{k},j,2}\end{array}
\right],\\
\tilde{c}_{\boldsymbol{k},j,\nu}=& \left[\begin{array}{c}
c_{\boldsymbol{k},j,\nu,\uparrow}\\
c_{\boldsymbol{k},j,\nu,\downarrow}\end{array}\right],\\
\xi_{TI} =&( M_0 - 2b_1 +2b_2 \cos(k_xa) + 2b_2\cos(k_ya) - 4b_2 -\mu_{TI})\delta_{j,j'}
+ b_1 (\delta_{j,j'+1} + \delta_{j,j'-1} )
, \\
\boldsymbol{A}=&(a_2k_x\delta_{j,j},a_2k_y\delta_{j,j},-ia_1(\delta_{j,j'+1}-\delta_{j,j'-1})),
\end{align}
where $c^\dagger_{\boldsymbol{k},j,\mu,s}$ ($c_{\boldsymbol{k},j,\nu,s}$) is the creation (annihilation) operator
of an electron with spin $s$, belonging to the band $\nu=1-2$, having two-dimensional wave vector $\boldsymbol{k}=(k_x,k_y)$, 
and at a lattice site $j< 0$ in the $z$ direction. We used the periodic boundary condition in the $xy$ plane.

In the same way, we describe the FI by
\begin{align}
H_{FI} =& \sum_{j,j'}\sum_{\boldsymbol{k}}
\left[
\tilde{c}^\dagger_{\boldsymbol{k},j',1},\tilde{c}^\dagger_{\boldsymbol{k},j',2}\right]
\left[\begin{array}{cc}
(\xi_{FI} +E_1)\hat{s}_0 & 0 \\
0 & (-\xi_{FI} +E_2)\hat{s}_0  + \boldsymbol{M}\cdot \hat{\boldsymbol{s}}
\end{array}\right]
\left[\begin{array}{c}
\tilde{c}_{\boldsymbol{k},j,1}\\\tilde{c}_{\boldsymbol{k},j,2}\end{array}
\right],\\
\xi_{FI} =& (2t \cos(k_x) + 2t\cos(k_y) - 8t -\mu_{FI})\delta_{j,j'} +t (\delta_{j,j'+1} + 
\delta_{j,j'-1} ), 
\end{align}
for $j>0$. At the interface ($j=0$), TI and FI are connected by,   
\begin{align}
H_{I}= &\sum_{\boldsymbol{k}} 
\left[
\tilde{c}^\dagger_{\boldsymbol{k},0,1},\tilde{c}^\dagger_{\boldsymbol{k},0,2}\right]
\left[\begin{array}{cc}
(\xi_{I} +E_1/2)\hat{s}_0 & \boldsymbol{A}'\cdot\hat{\boldsymbol{s}} \\
\boldsymbol{A}'\cdot\hat{\boldsymbol{s}}  & (-\xi_{I} +E_2/2)\hat{s}_0  + \boldsymbol{M}\cdot \hat{\boldsymbol{s}}/2
\end{array}\right]
\left[\begin{array}{c}
\tilde{c}_{\boldsymbol{k},0,1}\\\tilde{c}_{\boldsymbol{k},0,2}\end{array}
\right],\\
2\xi_{I} =& 
M_0 - 2b_1 +2(b_2+t) \cos(k_x) + 2(b_2+t)\cos(k_y) - 4b_2  - 8t
-\mu_{TI}-\mu_{FI},\\
2\boldsymbol{A}'=&(a_2k_x,a_2k_y,0).
\end{align}
\end{widetext}

\begin{figure*}[htbp]
 \begin{center}
  \subfigure{\includegraphics*[height=60mm]{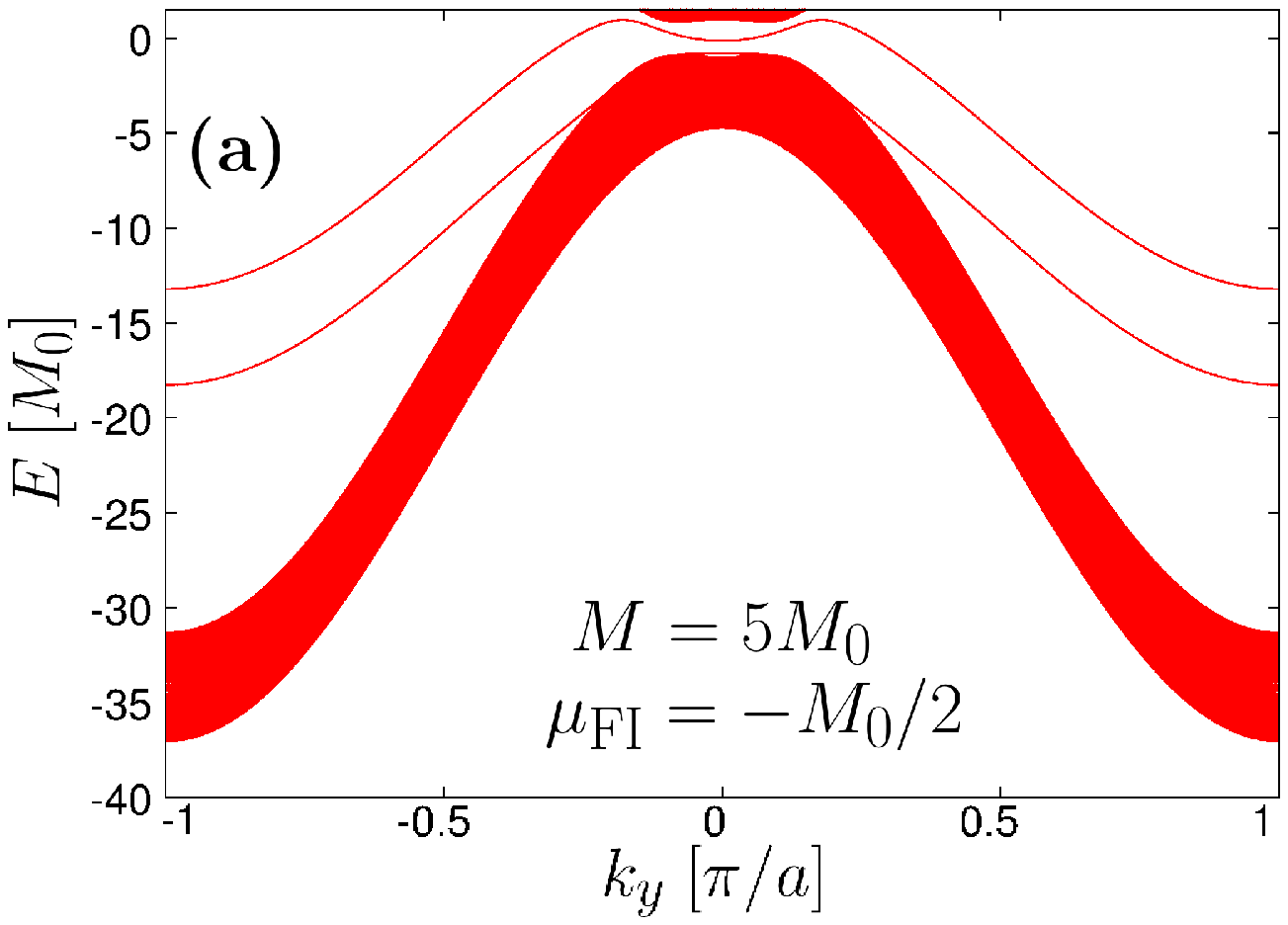}\label{fig:BAND_mzall}}
   \subfigure{\includegraphics*[height=60mm]{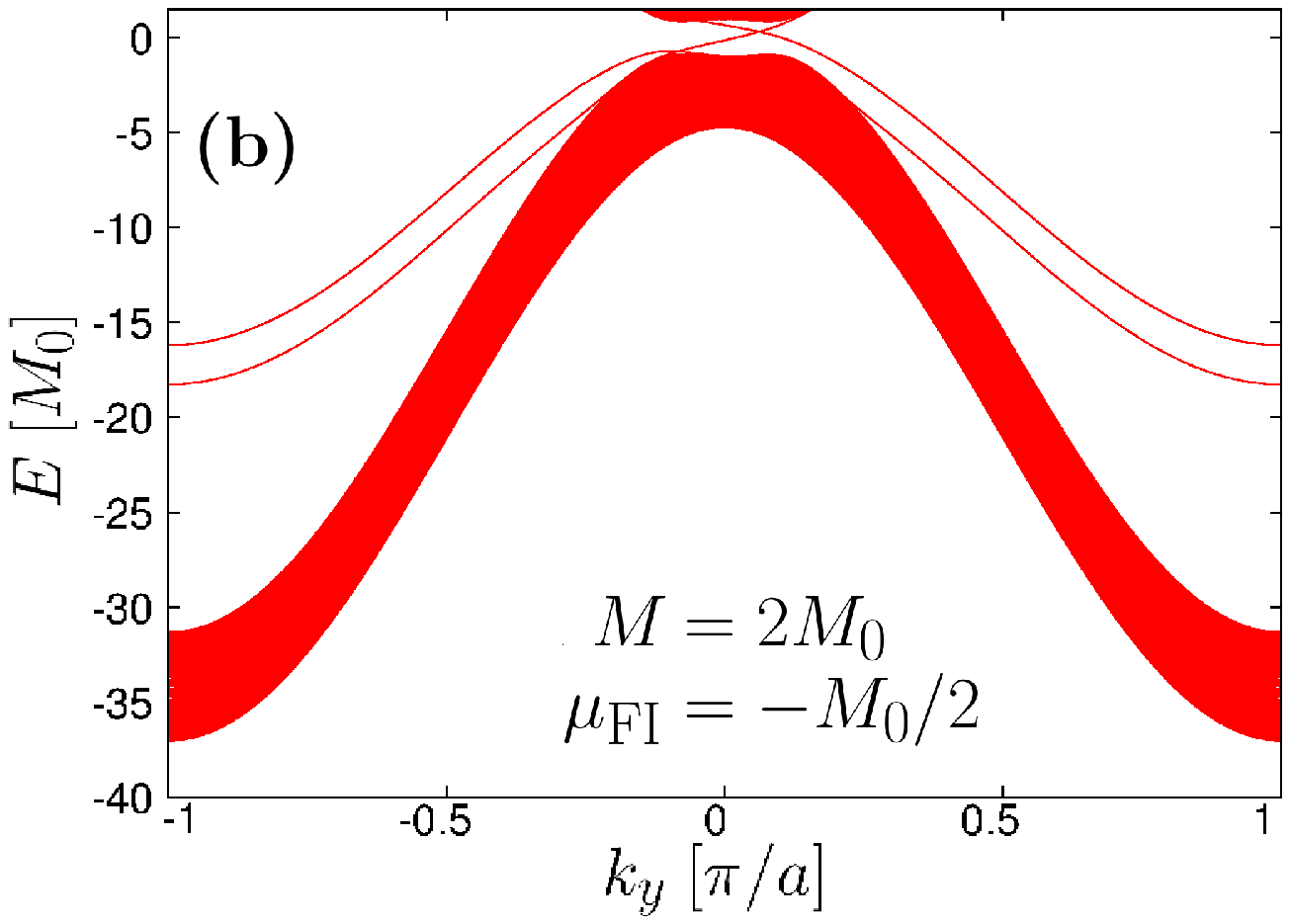}\label{fig:BAND_mxall}}
 \end{center}
\caption{The global pictures of band structures are showed for a perpendicular (a) and a parallel (b) magnetic moment.
 }
\end{figure*}

\begin{figure*}[htbp]
 \begin{center}
  \subfigure{\includegraphics*[height=60mm]{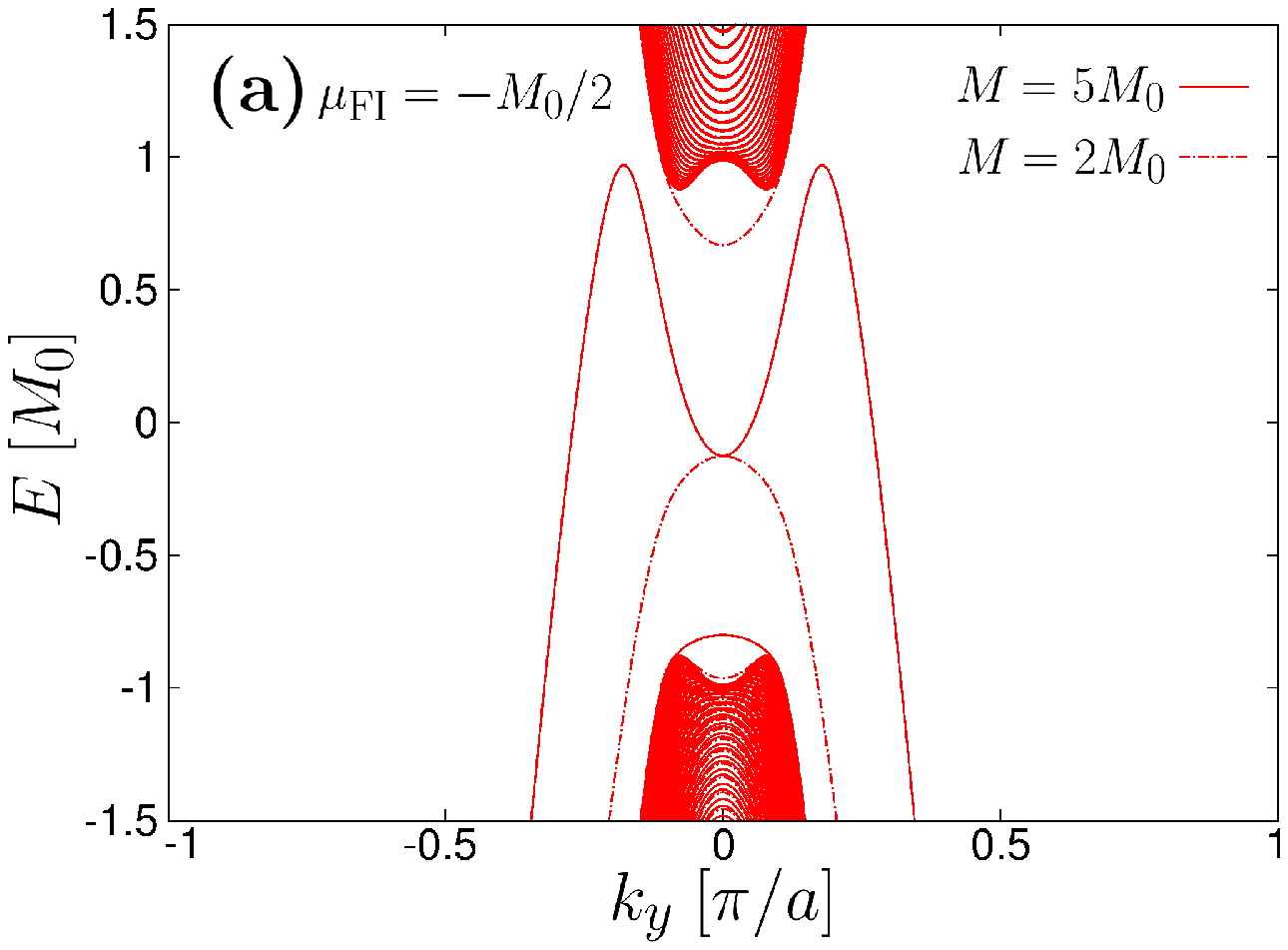}\label{fig:BAND_mz}}
   \subfigure{\includegraphics*[height=60mm]{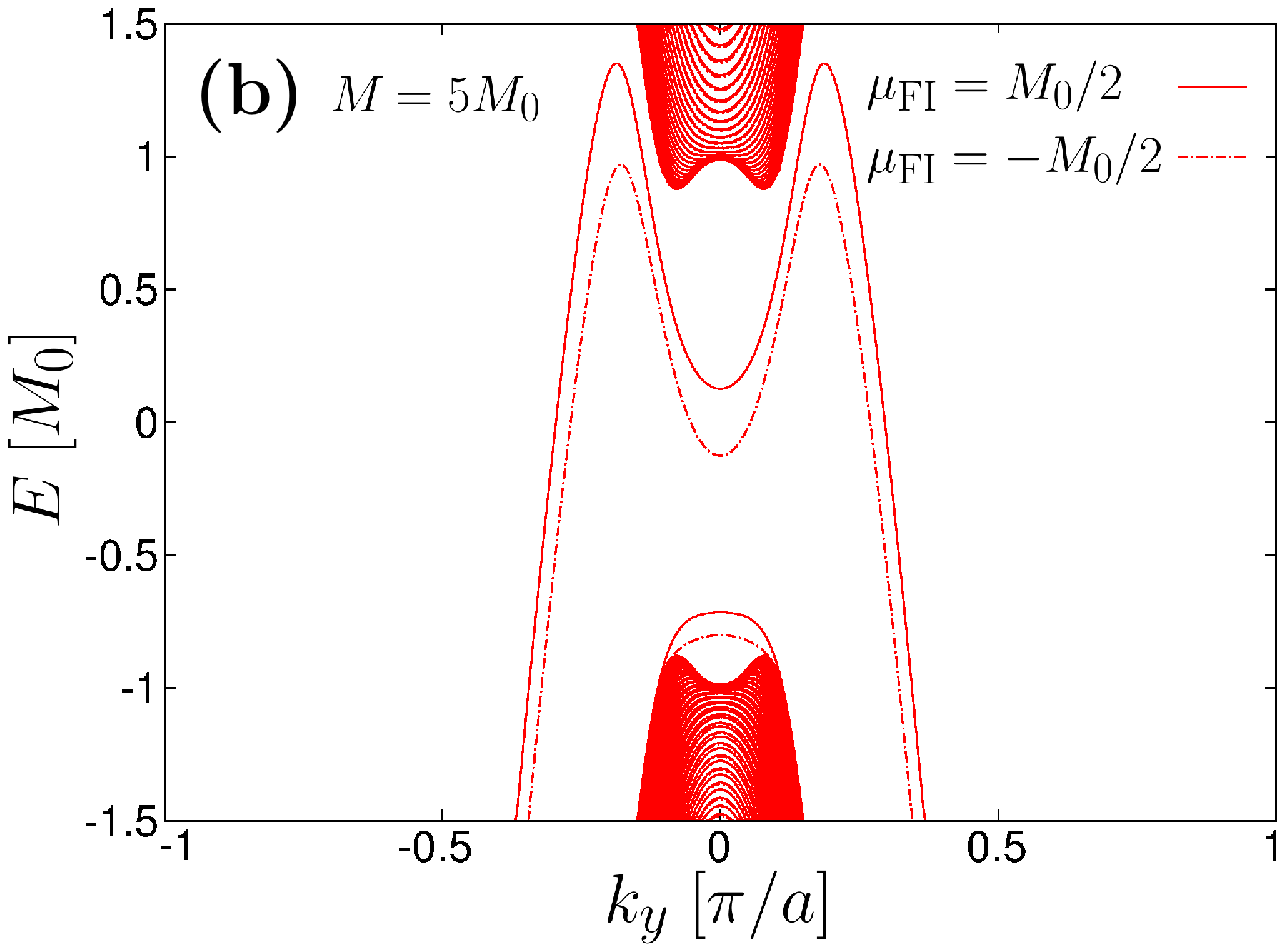}\label{fig:BAND_mzs}}
   \subfigure{\includegraphics*[height=60mm]{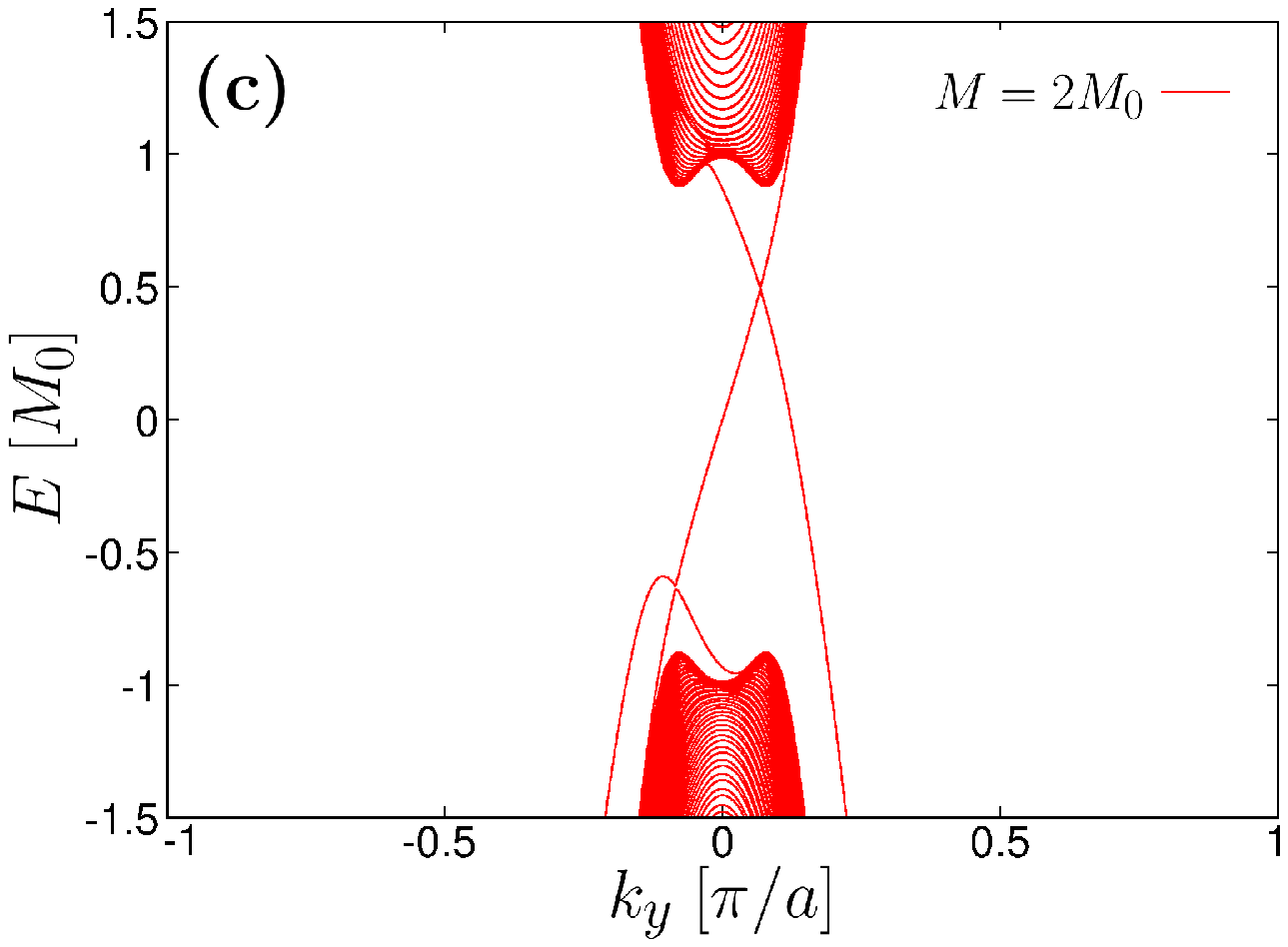}\label{fig:BAND_mx}}
   \subfigure{\includegraphics*[height=60mm]{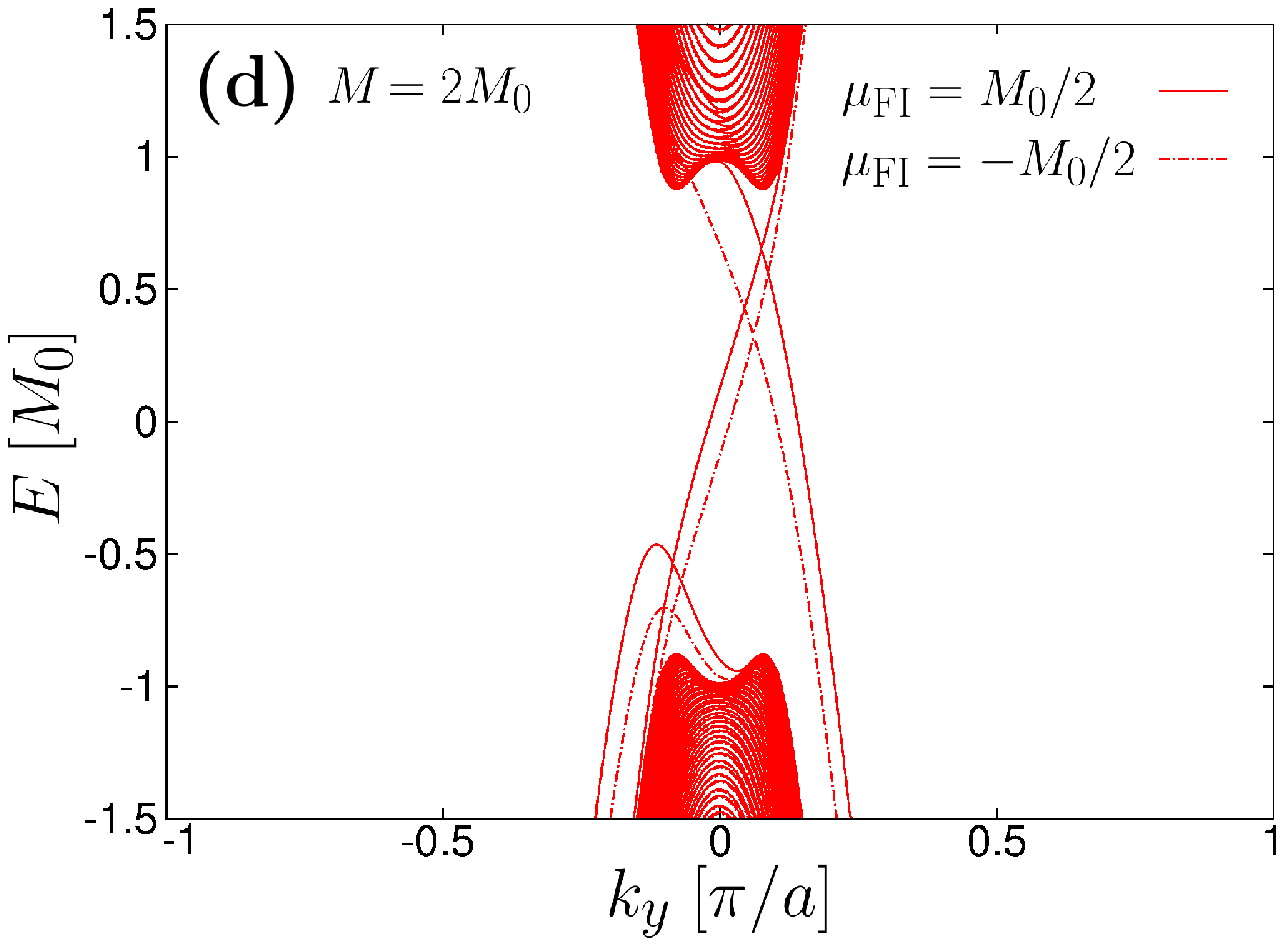}\label{fig:BAND_mxs}}
 \end{center}
\caption{The band structures of TI/FI junction with a perpendicular (a),(b) and parallel (c),(d) magnetic moment are plotted of the energy $E$ versus the wave vector $k_y$.
The optical gap of FI is locked in $2M_0$ in (a) and (c).
The magnitude of a magnetic moment of FI is $2M_0$ and $5M_0$.
There are the surface band separated from the bulk band structure.
The effect by shifting the Fermi energy in the optical gap is plotted in (b) and (d).
 }
\end{figure*}

The hard wall boundary condition along with $z$-axis is employed.
The parameters in this calculation take values of $\mathrm{Bi_2Se_3}$:  
$a_1=7.86M_0/a$, $a_2=14.6M_0/a$,  $b_1=3.57\times10M_0/a^2$, and $b_2=2.02\times10^2M_0/a^2$ in TI side\cite{H.Zhang2009}.
The lattice constant $a$ is about $5\; [\mathrm{\AA}]$.
In FI, we assume $b_{\mathrm{FI}}=10^{-2}b_1$ and $E_1=-E_2=-M/2$.
The total lattice size in the $z$ direction is 200 sites, where TI and FI occupy 150 and 50 sites, respectively.
A schematic band picture of a FI is shown in Fig.\ref{fig:FIscheme}.  
Electronic structure becomes asymmetric with respect to 
the Fermi level.

We first show the dispersion relations of the interface states rather large energy range 
for magnetic moment perpendicular to the interface (Fig.~\ref{fig:BAND_mzall}) and 
for magnetic moment parallel to the interface(Fig.~\ref{fig:BAND_mxall}), where 
the dispersion is calculated along $k_x=0$, $\mu_{FI}=-M_0/2$, $M=5M_0$ in (a) and $M=2M_0$ in (b).
The wave function of the interface states behaves like $e^{j/j_0}$ for $j<0$ in TI with $j_0$
being the inverse of localizing length. In the Figures, we also show the bulk band in TI.
As we discussed in Sec.~III, the upper dispersion in (a) is clearly separated from the bulk band of 
TI in whole Brillouin zone because of the band asymmetry in FI.
The dispersions of the interface states for the magnetic moment parallel to the interface 
have rather complicated structure as shown in (b). We note that upper dispersion branche 
is well separated from the bulk band for $|k_y| > 0.3$. We zoom up the dispersion relations near 
the $\Gamma$ point and discuss their features in the next figures.

In Fig.\ref{fig:BAND_mz}, we show the dispersion relation of the interface states
along $k_x=0$ for the magnetic moment perpendicular to the interface. 
Here we assume $\mu_{FI}=-M_0/2$ and show the results for $M=2M_0$ and $5M_0$.
When the magnetic moment is relatively small at $M=2M_0$, 
the Dirac cone disappears as predicted by the effective theory around the Dirac point. 
When we increase the exchange potential at $M=5M_0$, 
however, the dispersion of the interface state behaves like
 $\epsilon_{\boldsymbol{k}} \approx \alpha_0 - \alpha_2 k^2 + \alpha_4 k^4$. 
 As a result, the interface state become metallic.
Features of the metallic also depends on the fermi level in the FI.
The dispersion relation in Fig.~\ref{fig:BAND_mzs} show that 
the number of fermi surface is one for $\mu_{FI}=M_0/2$, whereas for $\mu_{FI}=-M_0/2$ two fermi surface appears.
These numerical results are consistent with analytical one's in Sec.~III.


Next we look into the interface states at TI/FI junction in the presence of the magnetic moment 
parallel to the junction plane.
Figure \ref{fig:BAND_mx} shows the dispersion relation along $k_x=0$
for $M\parallel x$, where $\mu_{FI}=0$ and $M=2M_0$.
There are two Dirac cones in the Brillouin zone, which is consistent with 
the argument in Sec.~III. 
In Fig.~\ref{fig:BAND_mxs}, we show the results at $M=2M_0$ for $\mu_{FI}=-M_0/2$ and 
$M_0/2$. The characteristic features of the interface states are
insensitive to parameters such as $\mu_{FI}$ and $M$.

\section{CONCLUSION}

In this paper, we have studied electronic properties of 
interface 
state between a 
topological insulator (TI) and a ferromagnetic insulator (FI) 
by using two-band model in three-dimension in both analytically 
and numerically. 
The energy gap of FI is usually much larger than that of TI and 
the band structures in FI is asymmetric with respect to 
its fermi level. 
The dispersion branches of the interface state are separated 
from the bulk band in whole Brillouin zone due to the asymmetry 
of the band structures. 
When the magnetic moment is in the perpendicular direction to the 
interface plane, the interface states become metallic. 
The number of fermi surfaces of such interface states depends 
on the material parameters. 
When the magnetic moment is in the parallel direction to the interface plane, 
metallic states always appear irrespective of the amplitude of 
the exchange potential.
The number of the Dirac point becomes even integers in whole Brillouin zone.
Such drastic effects of the magnetic moment on interface states obtained 
in analytical calculation have been confirmed by
the numerical simulation on the tight-binding lattice.

\section{acknowledgement}
This work was supported by 
the "Topological Quantum Phenomena" (No. 22103002) Grant-in Aid for 
Scientific Research on Innovative Areas from the Ministry of Education, 
Culture, Sports, Science and Technology (MEXT) of Japan.

\bibliography{TI-FIJ}

\end{document}